\documentclass{aa}  

\usepackage{graphicx}
\usepackage{txfonts}
\begin{document}

   \title{Mass - Effective Temperature - Surface Gravity Relation for Intermediate-Mass Main-Sequence Stars}

%   \subtitle{}

   \author{K{\i}l{\i}\c{c}o\u{g}lu, T.
          \inst{}
          }

   \institute{Ankara University, Faculty of Science, Department of Astronomy 
              and Space Sciences, 06100, Ankara, Turkey\\
              \email{tkilicoglu@ankara.edu.tr}
             }

   \date{Received September 15, 1996; accepted March 16, 1997}
   
   \titlerunning{MTGR for Intermediate-Mass MS Stars}

% \abstract{}{}{}{}{} 
% 5 {} token are mandatory
 
  \abstract
   {In this work, a mass - effective temperature - surface gravity relation (MTGR) is developed for main sequence stars in the range of $6400\,K \leqslant T_{\rm eff} \leqslant 20\,000\,K$ with log\,$g \gtrsim 3.44$. The MTGR allows the simple estimation of the masses of stars from their effective temperatures and surface gravities. It can be used for solar metallicity and can be rescaled for any metallicity within $-1.0 \leqslant [{\rm Fe/H}] \leqslant 0.7$. The effect of $\alpha$-enhanced compositions can also be considered with the help of correction terms. }
   {It is aimed to develop an MTGR that can estimate the masses of main-sequence stars from their atmospheric parameters. One advantage of an MTGR over the classical mass-luminosity relations is that its mass estimation is based on parameters that can be obtained by purely spectroscopic methods and, therefore, the interstellar extinction or reddening do not have to be known. The use of surface gravity $(g)$ also relates an MTGR with stellar evolution and provides a more reliable mass estimation.}
   {A synthetical MTGR is obtained from theoretical isochrones using a Levenberg-Marquardt $\chi^2$ minimization algorithm. The validity of the MTGR is then checked by testing over 278 binary components with precise absolute masses.}
   {Very good agreement has been obtained between the absolute masses of 278 binary star components and their masses estimated from the MTGR. A mathematical expression is also given to calculate the propagated uncertainties of the MTGR masses.}
   {For the typical uncertainties in atmospheric parameters and metallicity, i.e.,  $\pm2.8\%$ for $T_{\rm{eff}}$, $\pm0.1$\,dex for ${\rm log}\,g$, and $\pm0.15$\,dex for [Fe/H], the typical uncertainties in the masses estimated from the MTGR mostly remain around 5-9\%. The fact that this uncertainty level is only on average about three times as large as that of the absolute masses indicates that the MTGR is a very powerful tool for stellar mass estimation. A computer code, \texttt{mtgr.pro}, written in GDL/IDL is also provided for the relation.}

\keywords{stars: general -- stars: fundamental parameters -- (stars:) Hertzsprung-Russell and C-M diagrams -- (stars:) binaries: general -- stars: atmospheres -- stars: individual (AG Per, V1388 Ori, GG Lup)}

   \maketitle
%
%-------------------------------------------------------------------

\section{Introduction}
\label{sec:intro}

Mass is generally considered to be the most important fundamental parameter of a star for many applications in astrophysics. Mass, however, can be directly derived only for multiple systems. Asteroseismology and interferometry can be used to derive the mass (and also luminosity) of single stars with little model dependence \citep{cunhaetal07,whiteetal13}. The combination of these two techniques, however,  can only be applied to nearby stars of certain spectral types. Thanks to the mass-luminosity relations (MLRs), mass can be estimated for most single stars.  However, the luminosity of stars can not usually be obtained accurately from their apparent magnitude, as it requires good extinction information\footnote{ Spectral energy distributions (SEDs) may help to derive the level of extinction, but for many stars, it is not practically possible to construct an SED.} and a well-defined bolometric correction. The fact that MLRs are mostly affected by stellar evolution\footnote{ e.g., two stars of the same mass, one in zero-age-main-sequence and the other in terminal-age-main-sequence, may have different luminosities.} and metallicity also increases their uncertainty. Therefore, MLRs are not always the best solution for estimating the mass of single stars.

To avoid the effect of extinction, it can be useful to associate the mass with parameters purely derivable from spectroscopic methods, e.g., atmospheric parameters such as effective temperature ($T_{\rm{eff}}$) and surface gravity ($g$). There are many methods to derive these parameters using the optical spectra of stars, such as the ionization/excitation equilibrium of certain elements, modeling the hydrogen Balmer lines, and measuring the Balmer jump. The atmospheric parameters derived from these methods are basically extinction-free. The use of $g$ and [Fe/H] in these relations also allows stellar evolution and metallicity to be taken into account for the mass estimation. 

The masses of single stars can be estimated from their atmospheric parameters using $T_{\rm{eff}}$ - log\,$g$ diagrams with theoretical isochrones. This method requires the retrieval and reading of the isochrones with various ages and then the interpolation of isochrone parameters for the position of the targets. Although the application of the method is not difficult, it can be time-consuming when estimating the masses of a large number of stars. Moreover, the absence of a mathematical expression of this method makes it difficult to use in computer codes. A mass-effective temperature-surface gravity relation (MTGR) may thus allow us to estimate the masses of single stars more quickly and in a more code-friendly way.

Most theoretical and observational studies attempting to understand elemental abundance distributions in the atmospheres of stars are based on their mass \citep[see, e.g.,][]{turcotteetal98,richeretal00,kochukovandbagnulo06,dealetal18,monieretal19}. However, target stars in the observational studies may not always be members of binary systems or, even if they are, their components may not be bright enough to be detected. In this case, an MTGR may help to estimate the stellar masses.

An MTGR not only helps to estimate the mass of single stars, but can also be used for the elemental abundance analysis of double-lined spectroscopic binaries (SB2s), particularly for long-period ones that lack of accurate physical parameters. There are several groups working on deriving chemical abundances for binary stars \citep[e.g.,][]{folsometal12,gebranetal15,torresetal15}. Synthetic flux spectra of the stars can be obtained via various model atmosphere codes using $T_{\rm{eff}}$ and log\,$g$ primarily. However, these theoretical spectra usually simulate the unit surface area of the stars. To obtain the composite spectrum of an SB2 system, the synthetic flux spectra of its components must be scaled by the square of the radii and, therefore, the radius ratio of the system $R_1/R_2$ must be known. $R_1/R_2$ may not always be available or precise enough for long-period binaries, as long observation times are needed to cover a sufficient phase interval. Using the classical formula of gravity at a distance $R$ from the center of a mass $M$, $g = (GM/R^2)$ where G is the gravitational constant, one may easily see that the following relation can be written for detached binary stars with spherical components: \\

$\Big(\frac{\displaystyle R_1}{\displaystyle R_2}\Big)^2=\frac{\displaystyle M_1}{\displaystyle M_2}\cdot\frac{\displaystyle g_2}{\displaystyle g_1}$ \\

\noindent If $g_1$ and $g_2$ can be found spectroscopically, only the mass ratio is still needed to derive the squared radius ratio. An MTGR can be used at this point, as it can simply give a mass-ratio from the atmospheric parameters of the components.

In this paper, a mathematical MTGR model for main-sequence stars with $6400\,K\leqslant T_{\rm eff}\leqslant 20\,000\,K$ and log\,$g\gtrsim3.44$ is presented, which can be used in many different applications in astronomy similar to those listed above. Sec.\,\ref{sec:relation} explains how the relation is obtained with theoretical isochrones. In Sec\,\ref{sec:comp}, the validity of the MTGR is checked by testing over 278 binary components with precise absolute masses. Sec.\,\ref{sec:met} provides the MTGR solutions for non-solar metallicities. Sec.\,\ref{sec:alpha} gives correction terms for ${\rm \alpha}$-enhanced compositions. In Sec.\,\ref{sec:err}, a formulation of the propagated uncertainty of mass due to the uncertainties in $T_{\rm{eff}}$, log\,$g$ and [Fe/H] are given. A discussion and conclusion about the MTGR can finally be found in Sec.\,\ref{sec:disco}.

%--------------------------------------------------------------------
\section{Mass - effective temperature - surface gravity relation (MTGR) for $6400\,K \leqslant T_{\rm eff} \leqslant 20\,000\,K$ and log\,$g \gtrsim 3.44$}
\label{sec:relation}

A large number of stars with known absolute parameters ($M$, $T_{\rm{eff}}$, and log\,$g$) are needed for a reliable MTGR, as the problem is more complex than the two-parameter case, e.g., a mass-luminosity relation. The high-accuracy parameters of only several hundred detached binary components are available in the literature, and they are apparently insufficient for such a calibration. Because of this, $M$, $T_{\rm{eff}}$, and log\,$g$ parameters were initially collected from the theoretical isochrones for a large $T_{\rm{eff}}$ and log\,$g$ interval to derive a "synthetic" MTGR. 

The theoretical isochrones have been retrieved from \textsc{Parsec} \citep{bressanetal12}, mainly because it is well-maintained and includes many physical processes such as atomic diffusion and mass loss. Its web interface (\textsc{Cmd}) is also quite versatile for the development of such a relation. The {\small BaSTI}\footnote{\label{note}with overshooting and diffusion options} \citep{hidalgoetal18} and \textsc{Dartmouth}$^{\ref{note}}$ \citep{dotteretal08} evolutionary tracks/isochrones have also been used to determine the amount of systematic errors caused by differences in the assumptions of the theoretical models. A further comparison and discussion of many other theoretical evolutionary models used nowadays can be found in \citet{stancliffeetal16}. 

\begin{figure*}
\begin{center}
\includegraphics{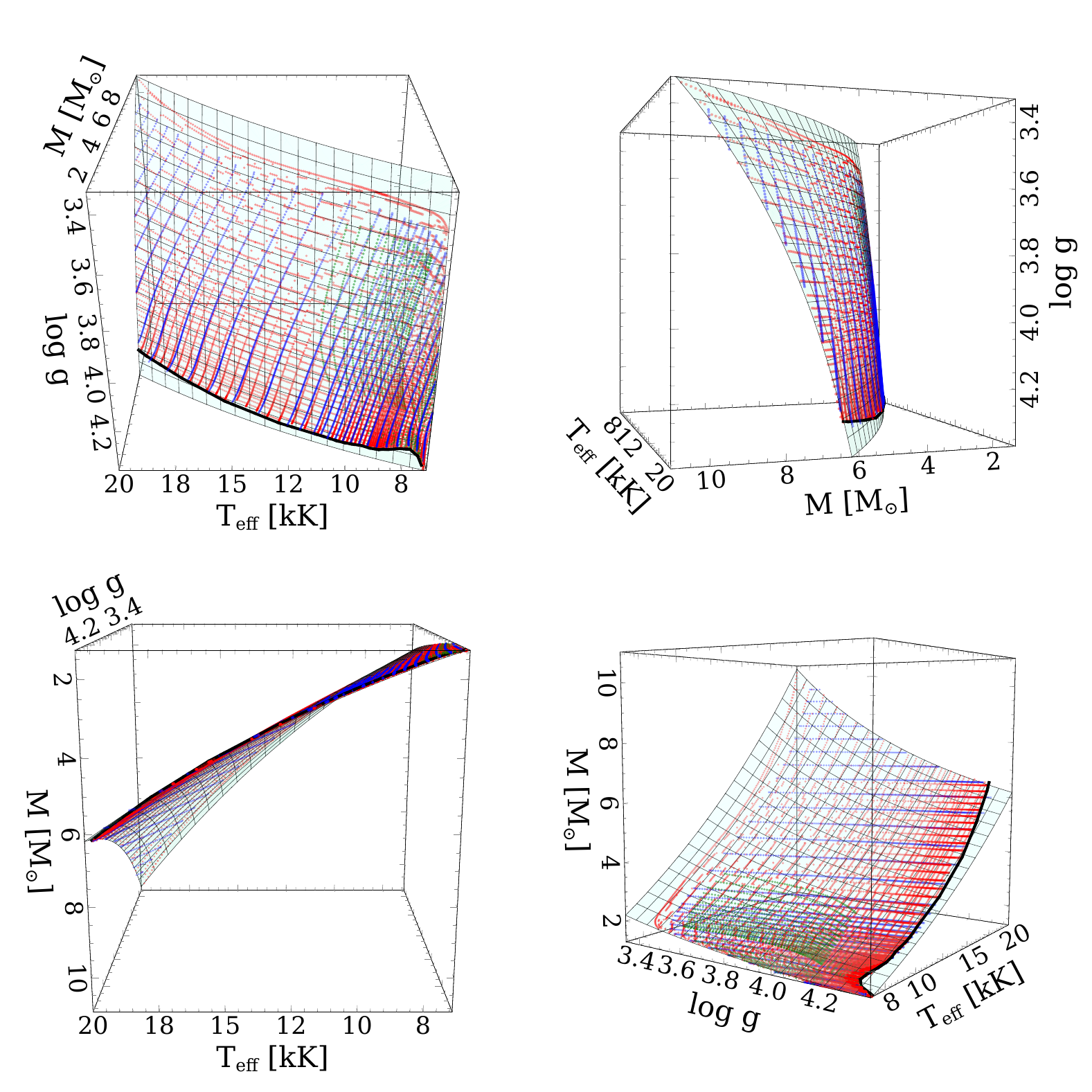}
\caption{$M$-$T_{\rm{eff}}$-log\,$g$ space from various view angles for main sequence stars with solar initial composition. The red, blue, and green circles are the points taken from \textsc{Parsec}, {\tiny BaSTI}, and \textsc{Dartmouth} isochrones/evolutionary tracks, respectively. The cyan surface with grids is the best surface fit to the theoretical \textsc{Parsec} isochrones. The surface corresponds to the synthetic MTGR formula (see text). The solid black line represents the approximate position of the ZAMS.}
\label{Fig1}
\end{center}
\end{figure*}

The isochrones were retrieved for $6.60\leqslant {\rm log}\,\tau \leqslant10.13$ with a step size of 0.01, where $\tau$ is the stellar age in years. The initial metallicity of the isochrones was set to $Z_{\rm ini}=0.01774$, which is the solar initial metallicity originally adopted by \citet{bressanetal12}. The theoretical $M$, $T_{\rm{eff}}$, and log\,$g$ trios were collected for the main-sequence phase (i.e., from zero-age-main-sequence [ZAMS] to near\footnote{For ${\rm log}g \lesssim 3.55$, there is a short-lived overall contraction phase in which the stars move downwards in the $T_{\rm{eff}}$-log\,$g$ diagram. This can result in two different masses from a single atmospheric parameter pair. Therefore, this contraction phase, which covers only about 1.5\% of the main sequence lifetime of stars, is not taken into account in the MTGR.} terminal-age-main-sequence [TAMS]) from the isochrone tables. I noticed that the isochrones form a smooth surface in $M$ - $T_{\rm{eff}}$ - log\,$g$ space between $6400\,K\,\leqslant\,T_{\rm eff}\,\leqslant\,20\,000\,K$ for main-sequence stars (red, blue, and green circles in Fig. \ref{Fig1}) and then chose this region for an MTGR development.

Using various mathematical functions, I attempted to obtain the most appropriate expression that relates these three observational variables. To derive the coefficients of these expressions, I fit the functions to the  $M$ - $T_{\rm{eff}}$ - log\,$g$ surface using a 2D version of Levenberg-Marquardt $\chi^2$ minimization routine \citep[see][ for its 1D version]{markwardt09}. I obtained the following expression for the MTGR that represents the theoretical $M$ - $T_{\rm{eff}}$ - log\,$g$ surface with good accuracy, i.e., the root-mean-square-deviation (RMSD) between the \textsc{Parsec} isochrones and the model is only 0.6\% for $M$: \\

$\bar{M} = a\bar{T} + b10^{c\bar{T}} + d\bar{g}10^{e\bar{g}}$ \\

\noindent where\footnote{all logarithms are to the base 10.}: \\

%$\bar{M}={\rm log}\,{\displaystyle \left(\frac{M}{M_{\odot}}\right)}$ \\
$\bar{M}={\rm log}\,{\displaystyle \left(M/M_{\odot}\right)}$

$\bar{T}={\rm log}\,T_{\rm{eff}}-4$

$\bar{g}={\rm log}\,g-4$ \\

\noindent and the coefficients are given by: \\

$a=+0.769687$

$b=+0.409173$

$c=+0.611016$

$d=-0.204335$

$e=-0.175664$ \\

The surface formed by the MTGR is shown in Fig.\,\ref{Fig1}. The RMSDs of the {\small BaSTI}  and \textsc{Dartmouth} tracks/isochrones from the surface are only 1.3\% and 1.9\% in the mass axis\footnote{e.g., the mass estimated using the \textsc{Parsec} models of a star with $T_{\rm{eff}}=10\,000$\,K and log\,$g=4.0$ is only 1.5\% larger than that estimated with the {\tiny BaSTI} models. }. Even for the most deviating points, the deviations are always less than 5\%. These deviations are in good agreement with the 1-2\% mass uncertainty reported by \citet{stancliffeetal16} for $3\,M_{\rm \odot}$ tracks taken from various stellar evolution codes.  These typical systematic errors of about 2\% in masses can be considered as a lower uncertainty limit of the MTGR.

As we later see in Sec. \ref{sec:comp}, the above synthetic MTGR, which is derived from theoretical isochrones, agrees well with the observations. The reader may, therefore, use this MTGR to estimate the mass of main-sequence stars for $6400\,K\,\leqslant\,T_{\rm eff}\,\leqslant\,20\,000\,K$, log\,$g\,\gtrsim\,3.44$, and solar-like composition. The computer code \texttt{mtgr.pro}\footnote{available from https://github.com/tolgahankilicoglu/mtgr or from the author upon request via e-mail.}, written in GDL\footnote{GNU Data Language \citep{coulaisetal10,coulais19}}/IDL\footnote{Interactive Data Language}, can also be used for this purpose. 

The MTGR can also be extrapolated for a  $T_{\rm eff}$ range down to 6200\,K and up to 25\,000\,K. However, it should be noted that for these extreme values, the uncertainties of the relation can be twice as large as those within the preferred $T_{\rm eff}$\,--\,log\,$g$ limits. For hot stars with $T_{\rm eff} \geqslant 15\,000\,{\rm K}$, the log\,$g$ limit of MTGR becomes smaller as TAMS shifts towards smaller log\,$g$ values, e.g., 3.38\,dex for 18\,000\,K and 3.33\,dex for 20\,000\,K.

\section{Comparison with observations}
\label{sec:comp}

The fundamental parameters of detached binary stars have recently been compiled by \citet{ekeretal18}. Their catalog consists of the data of 586 components studied in the years 1975-2017 \citep[see, e.g.,][]{griffinandgriffin09,zolaetal14,basturketal15,ozdarcanetal16,ratajczaketal16,torresetal17}. I have use only the data in the catalog with mass measurements accurate to $<15\%$. I have selected 278 stars in the catalog which fit the $T_{\rm eff}$, log\,$g$, and precision criteria of this study.

\begin{figure}
\begin{center}
\includegraphics[width=\columnwidth]{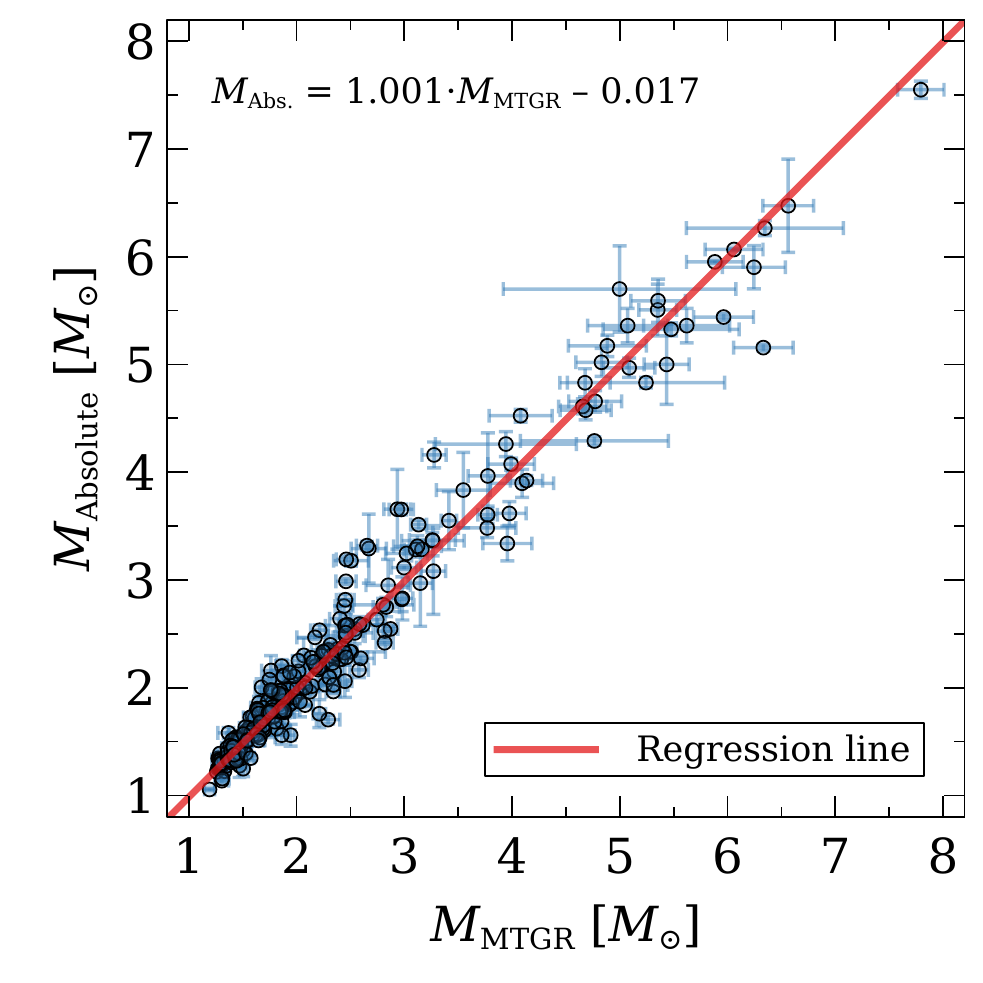}
\caption{Comparison of the absolute masses and the masses derived from the MTGR. Error bars in the y-axis are the uncertainties of the absolute masses given in \citet{ekeretal18}. Error bars in the x-axis are the propagated uncertainties for MTGR masses caused by the uncertainties in $T_{\rm eff}$ and log\,$g$ of the samples (see Sec. \ref{sec:err}). For the regression line, the points are weighted according to their measurement errors. }
\label{fig2}
\end{center}
\end{figure}

In Fig. \ref{fig2}, the absolute masses of these selected binary components have been compared to the masses derived from the MTGR with a regression line. Only four out of 278 objects showed remarkable deviations (more than two times the  root-mean-square-deviation) from the regression line. Two of these deviated objects are components of the same binary star, AG Per. The masses of its components have been given as $M_1=4.498 \pm 0.134\,M_{\odot}$ and $M_2=4.098 \pm 0.109\,M_{\odot}$ in the catalog of \citet{ekeretal18}. However, these masses do not match the original values, $M_1=5.36 \pm 0.16\,M_{\odot}$ and $M_2=4.90\pm0.13\,M_{\odot}$, given by \citet{gimenezandclausen94} and might be a typo in the catalog. The same inconsistency also appears in log\,$g$ values. I consequently used the masses and surface gravities given by \citet{gimenezandclausen94} for AG Per, and the new masses indeed follow the regression line in Fig. \ref{fig2} quite well. The third deviated object is the secondary component of V1388 Ori/HD 42401, whose primary component is hotter than the $T_{\rm eff}$ limit of this study and was not included. The mass of the secondary component has been derived as $M_2=5.16\pm0.13\,M_{\odot}$ by \citet{williams09}, who also reported that the star seems overluminous for its mass on the Hertzsprung-Russell Diagram (HRD). \citet{williams09} noted that the He I 4471 \AA\ line grows weaker while the Mg II 4481 \AA\ line becomes stronger as the temperature decreases and he has used these lines to estimate the effective temperature of the star as $T_{\rm eff}= 18\,500$\,K. However, in his Fig. 2, the synthetic profile of He I 4471 \AA\ is stronger and the synthetic profile of Mg II 4481 \AA\ is weaker than those observed for the secondary component. This indicates that the effective temperature of the secondary component might be overestimated in his study. A slightly reduced effective temperature ($\sim$16\,500-17\,000\,K) indeed reduces the luminosity of the star and places it in a more reasonable position in the HRD, i.e., close to the evolutionary track for $5\,M_{\odot}$. Using this reduced $T_{\rm eff}$, the MTGR also gives $M_2\approx5.3\,M_{\odot}$ for the star, which is very close to the absolute mass of $5.16\pm0.13\,M_{\odot}$ and makes the star very close to the regression line in Fig. \ref{fig2}. Nonetheless, I did not remove V1388 Ori B and used it as-is for the regression analysis. The fourth and final deviated star is the primary component of GG Lup. There is no plausible reason to exclude this star in Fig. \ref{fig2}, and it too was retained when deriving the slope of the regression line.

The slope of the final regression line in Fig. \ref{fig2} is 1.001 with an almost null zero-point offset\footnote{The small offset is most likely a result of the systematic errors discussed in sec. \ref{sec:relation}} ($-0.017$). This is good observational proof showing the robustness of the synthetic MTGR. . For the data and regression line, the RMSD is 0.56\,$M_{\odot}$ and the mean/median-absolute-deviation (MAD) is $\sim0.1\,M_{\odot}$. Here, the large value of RMSD arises not only from the approximations in MTGR (such as solar initial metallicity and no rotation), but also from the observational uncertainties in the $T_{\rm eff}$, log\,$g$, and $M_{\rm Abs.}$ parameters of the samples. RMSD, thus, may not reflect the typical uncertainty of the MTGR. The MAD may also be biased and underestimated because the cataloged binary star samples are mostly concentrated in the $M \lesssim 2.5\,M_{\odot}$ region, where the stars show less scatter in Fig. \ref{fig2}. By dividing the samples into two groups, I conclude that the typical uncertainties for the masses derived by the MTGR are about $\pm0.1\,M_{\odot}$ for $M \lesssim 2.5\,M_{\odot}$ and about $\pm0.3\,M_{\odot}$ for $M \gtrsim 2.5\,M_{\odot}$. This roughly corresponds to a $6-7\%$ typical uncertainty in mass for the cataloged binary star samples. For a detailed uncertainty calculation, see Sec.\,\ref{sec:err}.

\section{Effect of metallicity}
\label{sec:met}

One may need to constrain metallicity $\rm [M/H]$ or $Z$ in the MTGR. To see how the metallicity affects the MTGR, I repeated the same procedure as in Sec.\,\ref{sec:relation} for the isochrones with $\rm [M/H]=[-0.4, -0.3, -0.2, -0.1$, 0.0, 0.1, 0.2, 0.3, 0.4, 0.5, 0.6, 0.7] and derived the MTGR coefficients $a$, $b$, $c$, $d$, and $e$ for each. The corresponding ${Z}$ values of these metallicities has been given in the second column of Table \ref{tab:valid}. 
%\simeq [0.00758,  0.00847, 0.00947, 0.01058, 0.01182, 0.01319, 0.01471, 0.01640, 0.01826, 0.02032, 0.02259, 0.02510, 0.02784], respectively\footnote{${\rm [M/H]}={\rm log}\left( {\displaystyle \frac{N_{\rm (Z \geq 3), star}}{N_{\rm H, star}}}\right)-{\rm log}\left({\displaystyle \frac{N_{(Z \geq 3), \odot}}{N_{\rm H, \odot}}}\right)$}.
As the metallicity increases, the isochrone points shift to the lower right of the HRD and the validity range of the MTGR, therefore, changes slightly. Table \ref{tab:valid} gives the list of these validity ranges for various metallicities.

\begin{table}
\caption{Validity range of the MTGR for various metallicities} 
\centering
\begin{tabular}{llcc}
\hline\hline
$\rm{[M/H]}$ & Z & $T_{\rm eff}~[{\rm K}]$ range & log\,$g$ limit \\
\hline
$-0.4$ & $0.00605$ & $7040 - 22\,500$ & $\gtrsim 3.56$ \\
$-0.3$ & $0.00758$ & $6840 - 21\,700$ & $\gtrsim 3.53$ \\
$-0.2$ & $0.00947$ & $6770 - 21\,200$ & $\gtrsim 3.50$ \\
$-0.1$ & $0.01182$ & $6550 - 20\,400$ & $\gtrsim 3.47$ \\
$0.0$ & $0.01471$ & $6420 - 20\,200$ & $\gtrsim 3.44$ \\
$0.0865$* & 0.01774* & $6400 - 20\,000$ & $\gtrsim 3.44$ \\
$0.1$ & $0.01826$ & $6400 - 20\,000$ & $\gtrsim 3.44$ \\
$0.2$ & $0.02259$ & $6390 - 19\,000$ & $\gtrsim 3.41$ \\
$0.3$ & $0.02784$ & $6380 - 18\,500$ & $\gtrsim 3.39$ \\
$0.4$ & $0.03414$ & $6360 - 18\,300$ & $\gtrsim 3.37$ \\
$0.5$ & $0.04162$ & $6355 - 17\,800$ & $\gtrsim 3.34$ \\
$0.6$ & $0.05039$ & $6350 - 17\,300$ & $\gtrsim 3.34$ \\
$0.7$ & $0.06051$ & $6340 - 17\,000$ & $\gtrsim 3.33$ \\
\hline
\multicolumn{4}{l}{*\footnotesize Solar initial metallicity used for the  metallicity-independent} \\
\multicolumn{4}{l}{\ \ \footnotesize coefficients.}
\end{tabular} \label{tab:valid}
\end{table}

\begin{figure}
\begin{center}
\includegraphics[width=\columnwidth]{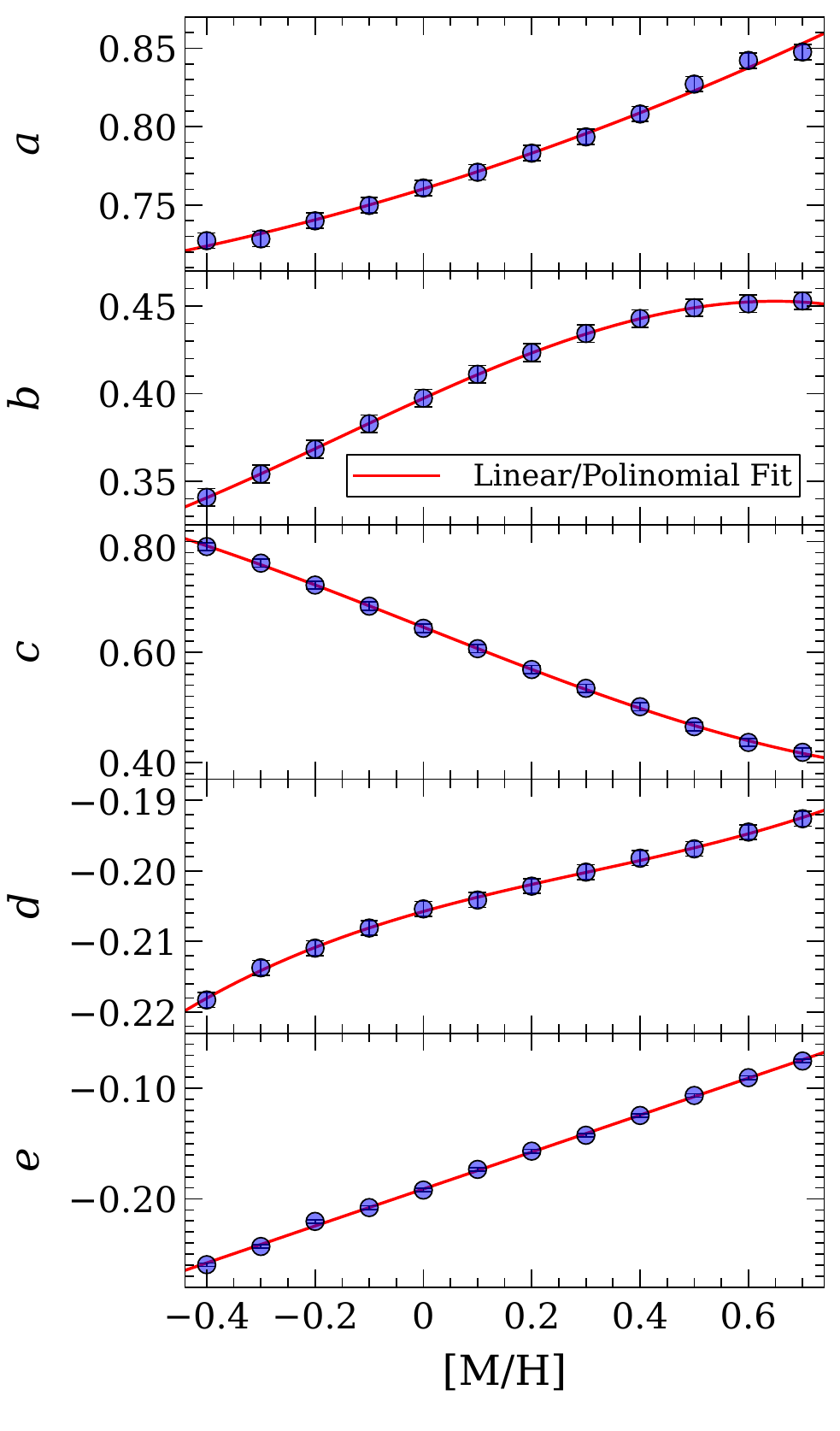}
\caption{Changes in the MTGR coefficients with metallicity. The points on the chart contain error bars.}
\label{fig3}
\end{center}
\end{figure}

Fig. \ref{fig3} shows the behavior of the MTGR coefficients for various metallicities. The coefficients change slightly quadratically/cubically as a function of metallicity, except for $e$, which exhibits a linear tendency. These relationships can be expressed as follows: \\

$a = +0.037539\,\mathrm{[M/H]}^{2}+0.106185\,\mathrm{[M/H]}+0.760313$

$b = -0.074620\,\mathrm{[M/H]}^{3}-0.035236\,\mathrm{[M/H]}^{2} +
\\ \null\hspace{1.34cm} 0.139705\,\mathrm{[M/H]}+0.397282$

$c = +0.122190\,\mathrm{[M/H]}^{3} +0.001445\,\mathrm{[M/H]}^{2} -
\\ \null\hspace{1.34cm} 0.387297\,\mathrm{[M/H]}+0.645036$

$d = +0.017391\,\mathrm{[M/H]}^{3} -0.015925\,\mathrm{[M/H]}^{2} +
\\ \null\hspace{1.34cm} 0.021622\,\mathrm{[M/H]}-0.205739$

$e = +0.167246\,\mathrm{[M/H]}-0.191036$ \\

\noindent where\footnote{${\rm log}{(\frac{Z}{Z_{\odot}})}$ does not have to be exactly equal to $[{\rm Fe/H}]$, because the metallicity $Z$ is mostly driven by $[{\rm C/H}]$ and, particularly, $[{\rm O/H}]$. However, [{\rm Fe/H]} can be preferred to scale $Z$ for scaled-solar compositions, because iron is an element showing large number of lines in the spectra of the stars and its abundance is generally much more reliable than those of carbon and oxygen. A more precise approximation formula, ${\rm [M/H] \simeq log}((Z/X)/0.0207)$, given by \citet{bressanetal12} used for the [M/H] values in Table \ref{tab:valid}.}: \\

$\rm{[M/H]} \simeq {\rm log}{\displaystyle \left( \frac{Z}{Z_{\odot}} \right)} \simeq [{\rm Fe/H}]$ \\

Calculating the coefficients above, the MTGR can be used for any metallicities in the range of $-0.4 \lesssim [{\rm Fe/H}] \lesssim 0.7$. Masses calculated with the MTGR are greater than $\sim$1 $M_{\odot}$ in the range where the relation is valid. Therefore, this paper mostly covers Population I stars. Although such stars are unlikely to have formed in an environment with ${\rm [Fe/H]} < -0.4$, I also present formulae for calculating MTGR coefficients for $-1.0 \leqslant {\rm [Fe/H]} < -0.4$ in Appendix \ref{lowmetal}.

I note that [Fe/H] here is the iron abundance\footnote{${\rm [Fe/H]}={\rm log}\left( {\displaystyle \frac{N_{\rm Fe, star}}{N_{\rm H, star}}}\right)-{\rm log}\left({\displaystyle \frac{N_{\rm Fe, \odot}}{N_{\rm H, \odot}}}\right)$} of the "whole" star and does not have to be equal to the photospheric [Fe/H]. Because of atomic diffusion, the photospheric elemental abundances of stars may vary during their evolution. These variations become more evident in stars with high effective temperatures and low rotational velocities \citep[see, e.g.,][]{talon06}. In order to use the MTGR for stars whose photospheric chemical abundances are thought to be strongly influenced by diffusion (or by any other physical process): i) if the object is a member of a stellar group (i.e., cluster/association/moving group), it would be best to scale the MTGR using the values of [Fe/H] or $Z$ proposed for that group or ii) if the object is a field star without any defined group membership, it may be more reasonable to use the original coefficients\footnote{corresponding to $Z=0.01774$ or $\rm{[M/H]}\simeq0.08653$} given in Sec. \ref{sec:relation} instead of scaling the MTGR with the photospheric [Fe/H]. The MTGR can also be used for chemically peculiar stars, but it should never be scaled by the [Fe/H] values derived for them, as the atmospheres of these unusual objects do not represent their interiors.

\section{Effect of $\alpha$-enhanced composition}
\label{sec:alpha}

Current $\alpha$-element abundance gradient studies for the galaxy confirm that stars can be born in environments with not only solar-like $\alpha$-element abundances but also $\alpha$-enhanced abundances \citep[see, e.g.,][]{gonzalezetal11,johnsonetal11,duongetal19,hayesetal20}.  Although the relationship between $\alpha$-element and iron abundances varies from region to region (i.e., thin disk, thick disc, halo, and bulge), ${\rm [\alpha/Fe]}$ generally tends to increase with decreasing [Fe/H] \citep[see, e.g.,][]{crestanietal21}. Since these differences in $\alpha$-element abundances can affect mass calculations, this section discusses how changes in $[{\rm \alpha/Fe}]$ can be accounted for in the MTGR. 

I have used the publicly available {\small BaSTI} scaled-solar and $\alpha$-enhanced \citep[][for ${\rm [\alpha/Fe]}$=+0.4]{pietrinfernietal21} stellar evolutionary tracks to characterize the effect of $\alpha$-enhanced compositions on the MTGR masses. In order to avoid systematic differences between the \textsc{Parsec} and {\small BaSTI} models or systematic errors due to the assumptions in the models as much as possible, I have used the ratio of mass changes with $\alpha$-element abundance instead of the mass values themselves. For this purpose, I define a $r(M_{\rm \alpha})$ ratio as follows: \\

$r(M_{\rm \alpha}) = \dfrac{M_{[{\rm \alpha/Fe}]=+0.4}}{M_{[{\rm \alpha/Fe}]=0.0}}-1$  \\

\noindent Here, $r(M_{\rm \alpha})$ represents the ratio of the masses obtained with the {\small BaSTI} models with $[{\rm \alpha/Fe}]=+0.4$ and $[{\rm \alpha/Fe}]=0.0$ for the same $T_{\rm eff}, {\rm log}\,g$, and [Fe/H] values.

\begin{figure}
\begin{center}
\includegraphics[width=\columnwidth]{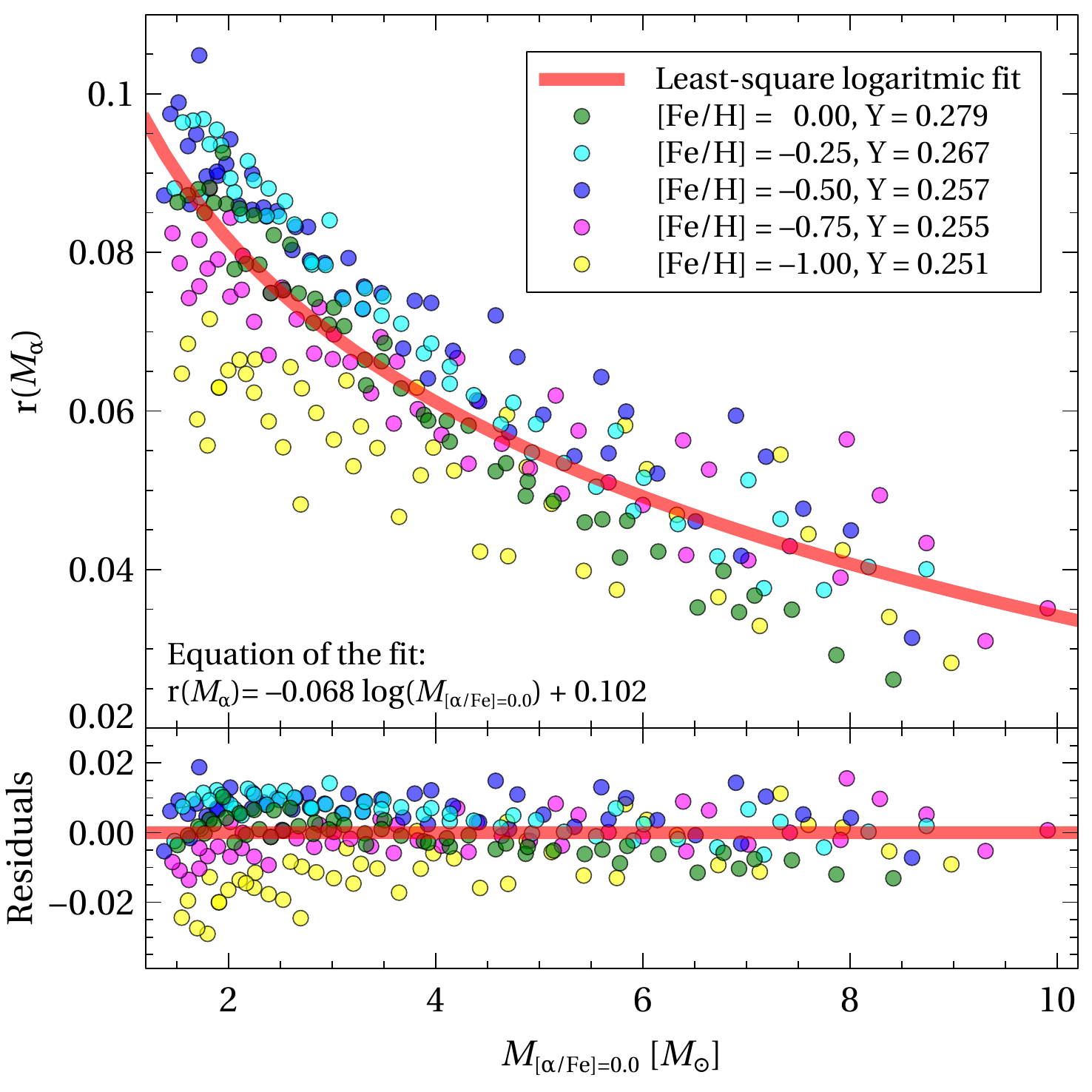}
\caption{Behavior of $r(M_{\rm \alpha})$ with $M_{[{\rm \alpha/Fe}]=0.0}$. Y is the mass fraction of helium used in $\alpha$-enhanced models. }
\label{fig4}
\end{center}
\end{figure}

In order to examine the behavior of $r(M_{\rm \alpha})$, I retrieved {\small BaSTI} evolutionary tracks in the range of 1 to 13 $M_{\odot}$ in steps of 0.01 $M_{\odot}$, for [Fe/H]=[$-1.00$, $-0.75$, $-0.50$, $-0.25$, 0.00], and also for $[{\rm \alpha/Fe}]=[0.0, +0.4]$. Fig. \ref{fig4} shows the calculated $r(M_{\rm \alpha})$ values for the evenly spaced combinations of various ${\rm log}\,T_{\rm{eff}}$, ${\rm log}\,g$, and [Fe/H] values (247 sets in total). 

It can be seen from Fig. \ref{fig4} that an increase of 0.4 dex in $[{\rm \alpha/Fe}]$ results in an increase of between 2\% and 11\% in mass. The magnitude of this effect decreases from low-mass stars to more massive ones. The $r(M_{\rm \alpha})$ values degenerate with the change of log\,$g$ and [Fe/H] values. It is seen that the degeneration due to log\,$g$ increases as [Fe/H] decreases. The RMSD of the scatter from the least-square logarithmic fit is 0.9\%, about half the typical systematic error of 2\% given in Sec. \ref{sec:relation} for the theoretical stellar evolution models. A correction term calculated from this scatterplot, therefore, can approximately reflect the effect of $\alpha$-element abundance changes on the MTGR.

By using the coefficients of the logarithmic fit, setting $M_{\rm MTGR} \cong M_{[{\rm \alpha/Fe}]=0.0}$, and linearly scaling $[{\rm \alpha/Fe}]$ abundances, the following expression was obtained: \\

$M_{\rm MTGR-\alpha} = M_{\rm MTGR}(1+{\rm [\alpha/Fe]}(k_{1}{\rm log}(M_{\rm MTGR})+k_{2}))$ \\

\noindent where: \\

$k_{1}=-0.17$

$k_{2}=0.255$ \\

\noindent By  placing the $\alpha$-element abundance (${\rm [\alpha/Fe]}$) suitable for the region where the star is located and the mass of the star calculated from the MTGR ($M_{\rm MTGR}$) in the expression above, one can estimate the corrected mass ($M_{\rm MTGR-\alpha}$) by taking into account ${\rm \alpha}$-enhancement.

\section{Estimation of mass uncertainties}
\label{sec:err}

The mathematical solution of the error propagation of the MTGR is quite complex because the partial derivations of the formula with respect to $T_{\rm eff}$ and log\,$g$ contain a large number of terms. In order to estimate the uncertainty in the mass in a more practical way, I have determined the amount of change in the mass by moving $T_{\rm eff}$, log\,$g$, [M/H], and $\rm [\alpha/Fe]$ in the MTGR relationship within the error limits. Assuming the uncertainties in $T_{\rm eff}$, log\,$g$, [M/H], and $\rm [\alpha/Fe]$ are independent, the total uncertainty of the mass ($\Delta M$) can be estimated from the quadrature sum formula below: \\

$\Delta M \simeq \sqrt{(\Delta M_{T_{\rm{eff}}})^2+(\Delta M_{{\rm log}g})^2+(\Delta M_{{\rm [M/H]}})^2+(\Delta M_{{\rm [\alpha/Fe]}})^2}$ \\

\noindent where, $\Delta M_{T_{\rm eff}}$, $\Delta M_{{\rm log}g}$, $\Delta M_{{\rm [M/H]}}$, and $\Delta M_{{\rm [\alpha/Fe]}}$ are the changes in the mass due to the uncertainty in $T_{\rm eff}$, log\,$g$, [M/H], and ${\rm [\alpha/Fe]}$ (i.e., $\Delta T_{\rm eff}$, $\Delta {\rm log}\,g$, $\Delta {\rm [M/H]}$, and $\Delta {\rm [\alpha/Fe]}$), respectively. 

 In order to obtain the expressions of $\Delta M_{T_{\rm eff}}$ and $\Delta M_{{\rm log}g}$, I first performed a multiple linear regression analysis using the MTGR with the metallicity-independent coefficients (see Fig. \ref{FigB1}). From the quadrature sum of the two expressions, I obtained the following formula for estimating the mass uncertainty: \\

$\Delta M \simeq (((c_{1} T_{\rm{0}}+c_{2} g_0+c_{3})\Delta T_{\rm eff})^2 + $

\ \ \ \ \ \ \ \ \ \ $((c_{4}T_{\rm{0}}+c_{5}g_{0}+c_{6})\Delta{\rm log}\,g)^2)^{1/2}$

\noindent where: \\

$T_{\rm{0}} = T_{\rm{eff}} - 13\,200\,\rm{K}$

$g_{0} = {\rm log}\,g - 3.8 $
\\

$c_{1}= +2.59049\cdot10^{-8}$

$c_{2}= -2.66221\cdot10^{-4}$

$c_{3}= +4.72477\cdot10^{-4}$

$c_{4}= +2.64599\cdot10^{-4}$

$c_{5}= -3.42804$

$c_{6}= +2.56787$ \\

I then performed a similar regression analysis using the MTGR and the metallicity-dependent coefficients (see Fig. \ref{FigB2}) to examine the effect of metallicity on uncertainties. As can be seen from graphs 6 and 12 in Fig. \ref{FigB2}, the effect of [M/H] on the slopes of [$\Delta T_{\rm eff}-\Delta M_{T_{\rm eff}}$] and [$\Delta {\rm log}\,g-\Delta M_{{\rm log}g}$] was negligible and was, therefore, not included in the calculations of $\Delta M_{T_{\rm eff}}$ and $\Delta M_{{\rm log}g}$. However, parameters $T_{\rm eff}$, log\,$g$, and [M/H] all changed the slope of [$\Delta {\rm [M/H]}-\Delta M_{{\rm [M/H]}}$] and included in the calculation of $\Delta M_{\rm [M/H]}$ (see graphs 16-18 in Fig. \ref{FigB2}). From the quadrature sum of the found $\Delta M_{T_{\rm eff}}$, $\Delta M_{{\rm log}g}$, and $\Delta M_{\rm [M/H]}$ expressions, I obtained the following formula for estimating the mass uncertainty: \\

$\Delta M \simeq (((c_{1} T_{\rm{0}}+c_{2} g_0+c_{3})\Delta T_{\rm eff})^2 + $

\ \ \ \ \ \ \ \ \ \ $((c_{4}T_{\rm{0}}+c_{5}g_{0}+c_{6})\Delta{\rm log}\,g)^2 + $

\ \ \ \ \ \ \ \ \ \ $((c_{7}T_{\rm{0}}+c_{8}g_{0}+c_{9}\rm{[M/H]+c_{10}})\Delta{\rm { [M/H]}})^2)^{1/2}$

\noindent where: \\

$T_{\rm{0}}$, $g_{0}$: same as in the previous expression
\\

$c_{1}= +2.69332\cdot10^{-8}$

$c_{2}= -2.68997\cdot10^{-4}$

$c_{3}= +4.67700\cdot10^{-4}$

$c_{4}= +2.67543\cdot10^{-4}$

$c_{5}= -3.64229$

$c_{6}= +2.58160$

$c_{7}= +7.08798\cdot10^{-5}$

$c_{8}= -0.906971$

$c_{9}= -0.969023$ (for the entire [M/H] range)

$c_{9}= -1.53968$ (a better estimate for $-0.1 \leqslant [{\rm M/H}] \leqslant 0.6$)

$c_{10}= +1.07926$ (for the entire [M/H] range)

$c_{10}= +1.14431$ (a better estimate for $-0.1 \leqslant [{\rm M/H}] \leqslant 0.6$) \\

\noindent Here, I presented two solutions for $c_{9}$ and $c_{10}$, because the behavior of [$\Delta {\rm [M/H]}-\Delta M_{{\rm [M/H]}}$] slopes with [M/H] was quite non-linear (see graph 18 in Fig. \ref{FigB2}). 

By placing the observational variables $(T_{\rm{eff}}$,\,${\rm log}\,g,{\rm [M/H]})$ and their uncertainties ($\Delta{}T_{\rm{eff}}$,\,$\Delta{}{\rm log}\,g,\Delta{\rm { [M/H]}}$) in the expressions above, one can estimate the uncertainty in the MTGR mass ($\Delta{}M$) of the star. An expanded version of the formula including uncertainties in ${\rm \alpha}$-element abundances ($\Delta[{\rm \alpha/Fe}]$) can be found in Appendix \ref{alphauncert}. 

\section{Discussion and conclusion}
\label{sec:disco}

The uncertainty of the masses obtained with the MTGR depends on the uncertainties of $T_{\rm{eff}}$ and ${\rm log}\,g$ by nature. The mean uncertainty of the absolute masses for the selected 278 binary components is about 2.4\%, and this percentage can be regarded as the typical uncertainty for the mass derivation of binary stars in the selected $T_{\rm{eff}}$ and ${\rm log}\,g$ range. This uncertainty is also of the same order as the systematic differences between different evolutionary models. When the masses of the same samples are calculated with the MTGR, the average mass uncertainty rises to only 3.8\%. Considering that only the $T_{\rm{eff}}$ and ${\rm log}\,g$ values of the stars are used in the MTGR, it is seen that the uncertainty level achieved is at a level that can compete with absolute mass uncertainties on average. However, one should consider here whether the atmospheric parameters obtained by spectroscopic methods are as sensitive as those obtained from simultaneous light and radial velocity curve solutions of binary stars. 

It is seen that for the 278 stars in the catalog, the mean uncertainty in the $T_{\rm{eff}}$ is 2.8\%, and the mean uncertainty in ${\rm log}\,g$ is 0.02 dex. The typical uncertainty of 2.8\% for $T_{\rm{eff}}$ agrees well with those obtained from spectroscopic studies today (e.g., $T_{\rm{eff}} = 7500 \pm 200, 11\,000 \pm 300, 18\,000 \pm 500$). However, the typical uncertainty of 0.02 dex for ${\rm log}\,g$ in binary star solutions is much more precise than the $\sim{}0.05-0.20$ dex uncertainties that can be achieved by spectroscopic methods \citep[see, e.g.,][]{kochukhovetal06,fossatietal11,gebranetal16,kilicogluetal18}. Recent chemical abundance studies with spectral data of different quality also indicate that the iron abundance [Fe/H] of stars can be derived with an uncertainty mostly in the range of 0.08-0.20 dex. If the typical uncertainties of $T_{\rm{eff}}$, ${\rm log}\,g$, and [Fe/H] are accepted as $\pm2.8\%$, $\pm0.1$ dex, and $\pm0.15$ dex for spectroscopic studies, the uncertainties in the masses obtained from the MTGR are never more than 10.1\% and mostly remain in the range of 5-9\%. The uncertainties calculated here are propagated uncertainties due to the measurement errors of the observational variables and do not include systematic errors that may arise from phenomena such as rotation, magnetic field, surface inhomogeneity and wind. However, the comparison between the absolute masses and calculated MTGR masses in Sec. \ref{sec:comp} contains both random and systematic errors, and 80\% of the stars appear to have a deviation of less than 10\% in mass. For this reason, it is thought that systematic errors do not generally exceed random errors except for some outliers. Although these comparisons show that uncertainties in masses estimated with the MTGR are about 3 times higher than those in absolute masses on average, the method still has a good sensitivity in estimating the masses of single stars. The uncertainties of the spectroscopically obtained ${\rm log}\,g$ values should be small in order to derive more accurate masses from the MTGR.

In this study, an MTGR was developed that allows the calculation of the masses of main sequence stars from their $T_{\rm{eff}}$, ${\rm log}\,g$, and metallicity for $6400\,{\rm K} \lesssim T_{\rm{eff}} \lesssim 20\,000\,{\rm K}$. An expression was also given to calculate the uncertainty of the masses derived by the relation. The mean mass uncertainty was found to be 5-9\% for the typical $T_{\rm{eff}}$, ${\rm log}\,g$, and [M/H] uncertainties that can be achieved today by analyzing stellar spectra. In order to verify and improve the accuracy of the relation, there is a great need for studies of detached binary stars with well-defined chemical abundances (particularly for $\rm [\alpha/Fe]$ and $\rm [Fe/H]$). Further development of the relation for a wider range of $T_{\rm{eff}}$ and ${\rm log}\,g$ values may play a key role in the near future in estimating the masses of stars for spectroscopic space missions.

\begin{acknowledgements}
      The author would like to thank an anonymous referee for the valuable suggestions, which significantly improved the validity range of the relation and quality of this paper. The author would also like to thank S.O. Selam, H.V. \c{S}enavc\i{}, \"{O}. Ba\c{s}t\"{u}rk, and \c{S}. \c{C}al\i skan (Ankara University) for their useful comments.
\end{acknowledgements}


\bibliographystyle{aa}
\bibliography{kilicoglu.bib}

\begin{thebibliography}{37}
\expandafter\ifx\csname natexlab\endcsname\relax\def\natexlab#1{#1}\fi

\bibitem[{{Ba{\textcommabelow s}t{\"u}rk} {et~al.}(2015){Ba{\textcommabelow
  s}t{\"u}rk}, {Zola}, {Liakos}, {Nelson}, {Gazeas}, {{\"O}zavc{\i}},
  {Y{\i}lmaz}, {{\c{S}}enavc{\i}}, \& {Zakrzewski}}]{basturketal15}
{Ba{\textcommabelow s}t{\"u}rk}, {\"O}., {Zola}, S., {Liakos}, A., {et~al.}
  2015, \na, 41, 42

\bibitem[{{Bressan} {et~al.}(2012){Bressan}, {Marigo}, {Girardi}, {Salasnich},
  {Dal Cero}, {Rubele}, \& {Nanni}}]{bressanetal12}
{Bressan}, A., {Marigo}, P., {Girardi}, L., {et~al.} 2012, \mnras, 427, 127

\bibitem[{{Coulais}(2019)}]{coulais19}
{Coulais}, A. 2019, Astronomical Society of the Pacific Conference Series, Vol.
  523, {GDL - GNU Data Language 0.9.9}, ed. P.~J. {Teuben}, M.~W. {Pound},
  B.~A. {Thomas}, \& E.~M. {Warner}, 365

\bibitem[{{Coulais} {et~al.}(2010){Coulais}, {Schellens}, {Gales}, {Arabas},
  {Boquien}, {Chanial}, {Messmer}, {Fillmore}, {Poplawski}, {Maret}, {Marchal},
  {Galmiche}, \& {Mermet}}]{coulaisetal10}
{Coulais}, A., {Schellens}, M., {Gales}, J., {et~al.} 2010, Astronomical
  Society of the Pacific Conference Series, Vol. 434, {Status of GDL - GNU Data
  Language}, ed. Y.~{Mizumoto}, K.~I. {Morita}, \& M.~{Ohishi}, 187

\bibitem[{{Crestani} {et~al.}(2021){Crestani}, {Braga}, {Fabrizio}, {Bono},
  {Sneden}, {Preston}, {Ferraro}, {Iannicola}, {Nonino}, {Fiorentino},
  {Th{\'e}venin}, {Lemasle}, {Prudil}, {Alves-Brito}, {Altavilla}, {Chaboyer},
  {Dall'Ora}, {D'Orazi}, {Gilligan}, {Grebel}, {Koch-Hansen}, {Lala},
  {Marengo}, {Marinoni}, {Marrese}, {Mart{\'\i}nez-V{\'a}zquez}, {Matsunaga},
  {Monelli}, {Mullen}, {Neeley}, {da Silva}, {Stetson}, {Salaris}, {Storm},
  {Valenti}, \& {Zoccali}}]{crestanietal21}
{Crestani}, J., {Braga}, V.~F., {Fabrizio}, M., {et~al.} 2021, \apj, 914, 10

\bibitem[{{Cunha} {et~al.}(2007){Cunha}, {Aerts}, {Christensen-Dalsgaard},
  {Baglin}, {Bigot}, {Brown}, {Catala}, {Creevey}, {Domiciano de Souza},
  {Eggenberger}, {Garcia}, {Grundahl}, {Kervella}, {Kurtz}, {Mathias},
  {Miglio}, {Monteiro}, {Perrin}, {Pijpers}, {Pourbaix}, {Quirrenbach},
  {Rousselet-Perraut}, {Teixeira}, {Th{\'e}venin}, \& {Thompson}}]{cunhaetal07}
{Cunha}, M.~S., {Aerts}, C., {Christensen-Dalsgaard}, J., {et~al.} 2007, \aapr,
  14, 217

\bibitem[{{Deal} {et~al.}(2018){Deal}, {Alecian}, {Lebreton}, {Goupil},
  {Marques}, {LeBlanc}, {Morel}, \& {Pichon}}]{dealetal18}
{Deal}, M., {Alecian}, G., {Lebreton}, Y., {et~al.} 2018, \aap, 618, A10

\bibitem[{{Dotter} {et~al.}(2008){Dotter}, {Chaboyer}, {Jevremovi{\'c}},
  {Kostov}, {Baron}, \& {Ferguson}}]{dotteretal08}
{Dotter}, A., {Chaboyer}, B., {Jevremovi{\'c}}, D., {et~al.} 2008, \apjs, 178,
  89

\bibitem[{{Duong} {et~al.}(2019){Duong}, {Asplund}, {Nataf}, {Freeman}, {Ness},
  \& {Howes}}]{duongetal19}
{Duong}, L., {Asplund}, M., {Nataf}, D.~M., {et~al.} 2019, \mnras, 486, 3586

\bibitem[{{Eker} {et~al.}(2018){Eker}, {Bak{\i}{\textcommabelow s}}, {Bilir},
  {Soydugan}, {Steer}, {Soydugan}, {Bak{\i}{\textcommabelow s}},
  {Ali{\c{c}}avu{\textcommabelow s}}, {Aslan}, \& {Alpsoy}}]{ekeretal18}
{Eker}, Z., {Bak{\i}{\textcommabelow s}}, V., {Bilir}, S., {et~al.} 2018,
  \mnras, 479, 5491

\bibitem[{{Folsom} {et~al.}(2012){Folsom}, {Bagnulo}, {Wade}, {Alecian},
  {Landstreet}, {Marsden}, \& {Waite}}]{folsometal12}
{Folsom}, C.~P., {Bagnulo}, S., {Wade}, G.~A., {et~al.} 2012, \mnras, 422, 2072

\bibitem[{{Fossati} {et~al.}(2011){Fossati}, {Ryabchikova}, {Shulyak},
  {Haswell}, {Elmasli}, {Pandey}, {Barnes}, \& {Zwintz}}]{fossatietal11}
{Fossati}, L., {Ryabchikova}, T., {Shulyak}, D.~V., {et~al.} 2011, \mnras, 417,
  495

\bibitem[{{Gebran} {et~al.}(2016){Gebran}, {Farah}, {Paletou}, {Monier}, \&
  {Watson}}]{gebranetal16}
{Gebran}, M., {Farah}, W., {Paletou}, F., {Monier}, R., \& {Watson}, V. 2016,
  \aap, 589, A83

\bibitem[{{Gebran} {et~al.}(2015){Gebran}, {Hadrava}, {Jasniewicz}, \&
  {Richard}}]{gebranetal15}
{Gebran}, M., {Hadrava}, P., {Jasniewicz}, G., \& {Richard}, O. 2015, \apss,
  357, 137

\bibitem[{{Gimenez} \& {Clausen}(1994)}]{gimenezandclausen94}
{Gimenez}, A. \& {Clausen}, J.~V. 1994, \aap, 291, 795

\bibitem[{{Gonzalez} {et~al.}(2011){Gonzalez}, {Rejkuba}, {Zoccali}, {Hill},
  {Battaglia}, {Babusiaux}, {Minniti}, {Barbuy}, {Alves-Brito}, {Renzini},
  {Gomez}, \& {Ortolani}}]{gonzalezetal11}
{Gonzalez}, O.~A., {Rejkuba}, M., {Zoccali}, M., {et~al.} 2011, \aap, 530, A54

\bibitem[{{Griffin} \& {Griffin}(2009)}]{griffinandgriffin09}
{Griffin}, R.~E.~M. \& {Griffin}, R.~F. 2009, \mnras, 394, 1393

\bibitem[{{Hayes} {et~al.}(2020){Hayes}, {Majewski}, {Hasselquist}, {Anguiano},
  {Shetrone}, {Law}, {Schiavon}, {Cunha}, {Smith}, {Beaton}, {Price-Whelan},
  {Allende Prieto}, {Battaglia}, {Bizyaev}, {Brownstein}, {Cohen},
  {Frinchaboy}, {Garc{\'\i}a-Hern{\'a}ndez}, {Lacerna}, {Lane},
  {M{\'e}sz{\'a}ros}, {Bidin}, {M{\~{u}}noz}, {Nidever}, {Oravetz}, {Oravetz},
  {Pan}, {Roman-Lopes}, {Sobeck}, \& {Stringfellow}}]{hayesetal20}
{Hayes}, C.~R., {Majewski}, S.~R., {Hasselquist}, S., {et~al.} 2020, \apj, 889,
  63

\bibitem[{{Hidalgo} {et~al.}(2018){Hidalgo}, {Pietrinferni}, {Cassisi},
  {Salaris}, {Mucciarelli}, {Savino}, {Aparicio}, {Silva Aguirre}, \&
  {Verma}}]{hidalgoetal18}
{Hidalgo}, S.~L., {Pietrinferni}, A., {Cassisi}, S., {et~al.} 2018, \apj, 856,
  125

\bibitem[{{Johnson} {et~al.}(2011){Johnson}, {Rich}, {Fulbright}, {Valenti}, \&
  {McWilliam}}]{johnsonetal11}
{Johnson}, C.~I., {Rich}, R.~M., {Fulbright}, J.~P., {Valenti}, E., \&
  {McWilliam}, A. 2011, \apj, 732, 108

\bibitem[{{K{\i}l{\i}{\c{c}}o{\u{g}}lu}
  {et~al.}(2018){K{\i}l{\i}{\c{c}}o{\u{g}}lu}, {{\c{C}}al{\i}{\textcommabelow
  s}kan}, \& {{\"U}nal}}]{kilicogluetal18}
{K{\i}l{\i}{\c{c}}o{\u{g}}lu}, T., {{\c{C}}al{\i}{\textcommabelow s}kan},
  {\c{S}}., \& {{\"U}nal}, K. 2018, \apj, 852, 116

\bibitem[{{Kochukhov} \& {Bagnulo}(2006)}]{kochukovandbagnulo06}
{Kochukhov}, O. \& {Bagnulo}, S. 2006, \aap, 450, 763

\bibitem[{{Kochukhov} {et~al.}(2006){Kochukhov}, {Tsymbal}, {Ryabchikova},
  {Makaganyk}, \& {Bagnulo}}]{kochukhovetal06}
{Kochukhov}, O., {Tsymbal}, V., {Ryabchikova}, T., {Makaganyk}, V., \&
  {Bagnulo}, S. 2006, \aap, 460, 831

\bibitem[{{Markwardt}(2009)}]{markwardt09}
{Markwardt}, C.~B. 2009, in Astronomical Society of the Pacific Conference
  Series, Vol. 411, Astronomical Data Analysis Software and Systems XVIII, ed.
  D.~A. {Bohlender}, D.~{Durand}, \& P.~{Dowler}, 251

\bibitem[{{Monier} {et~al.}(2019){Monier}, {Griffin}, {Gebran},
  {K{\i}l{\i}{\c{c}}o{\u{g}}lu}, {Merle}, \& {Royer}}]{monieretal19}
{Monier}, R., {Griffin}, E., {Gebran}, M., {et~al.} 2019, \aj, 158, 157

\bibitem[{{{\"O}zdarcan} {et~al.}(2016){{\"O}zdarcan}, {{\c{C}}ak{\i}rl{\i}},
  \& {Akan}}]{ozdarcanetal16}
{{\"O}zdarcan}, O., {{\c{C}}ak{\i}rl{\i}}, {\"O}., \& {Akan}, C. 2016, \na, 46,
  47

\bibitem[{{Pietrinferni} {et~al.}(2021){Pietrinferni}, {Hidalgo}, {Cassisi},
  {Salaris}, {Savino}, {Mucciarelli}, {Verma}, {Silva Aguirre}, {Aparicio}, \&
  {Ferguson}}]{pietrinfernietal21}
{Pietrinferni}, A., {Hidalgo}, S., {Cassisi}, S., {et~al.} 2021, \apj, 908, 102

\bibitem[{{Ratajczak} {et~al.}(2016){Ratajczak}, {He{\l}miniak}, {Konacki},
  {Smith}, {Koz{\l}owski}, {Espinoza}, {Jord{\'a}n}, {Brahm}, {Hempel},
  {Anderson}, \& {Hellier}}]{ratajczaketal16}
{Ratajczak}, M., {He{\l}miniak}, K.~G., {Konacki}, M., {et~al.} 2016, \mnras,
  461, 2234

\bibitem[{{Richer} {et~al.}(2000){Richer}, {Michaud}, \&
  {Turcotte}}]{richeretal00}
{Richer}, J., {Michaud}, G., \& {Turcotte}, S. 2000, \apj, 529, 338

\bibitem[{{Stancliffe} {et~al.}(2016){Stancliffe}, {Fossati}, {Passy}, \&
  {Schneider}}]{stancliffeetal16}
{Stancliffe}, R.~J., {Fossati}, L., {Passy}, J.-C., \& {Schneider}, F.~R.~N.
  2016, \aap, 586, A119

\bibitem[{{Talon} {et~al.}(2006){Talon}, {Richard}, \& {Michaud}}]{talon06}
{Talon}, S., {Richard}, O., \& {Michaud}, G. 2006, \apj, 645, 634

\bibitem[{{Torres} {et~al.}(2017){Torres}, {McGruder}, {Siverd}, {Rodriguez},
  {Pepper}, {Stevens}, {Stassun}, {Lund}, \& {James}}]{torresetal17}
{Torres}, G., {McGruder}, C.~D., {Siverd}, R.~J., {et~al.} 2017, \apj, 836, 177

\bibitem[{{Torres} {et~al.}(2015){Torres}, {Sandberg Lacy}, {Pavlovski},
  {Fekel}, \& {Muterspaugh}}]{torresetal15}
{Torres}, G., {Sandberg Lacy}, C.~H., {Pavlovski}, K., {Fekel}, F.~C., \&
  {Muterspaugh}, M.~W. 2015, \aj, 150, 154

\bibitem[{{Turcotte} {et~al.}(1998){Turcotte}, {Richer}, \&
  {Michaud}}]{turcotteetal98}
{Turcotte}, S., {Richer}, J., \& {Michaud}, G. 1998, \apj, 504, 559

\bibitem[{{White} {et~al.}(2013){White}, {Huber}, {Maestro}, {Bedding},
  {Ireland}, {Baron}, {Boyajian}, {Che}, {Monnier}, {Pope}, {Roettenbacher},
  {Stello}, {Tuthill}, {Farrington}, {Goldfinger}, {McAlister}, {Schaefer},
  {Sturmann}, {Sturmann}, {ten Brummelaar}, \& {Turner}}]{whiteetal13}
{White}, T.~R., {Huber}, D., {Maestro}, V., {et~al.} 2013, \mnras, 433, 1262

\bibitem[{{Williams}(2009)}]{williams09}
{Williams}, S.~J. 2009, \aj, 137, 3222

\bibitem[{{Zola} {et~al.}(2014){Zola}, {{\c{S}}enavc{\i}}, {Liakos}, {Nelson},
  \& {Zakrzewski}}]{zolaetal14}
{Zola}, S., {{\c{S}}enavc{\i}}, H.~V., {Liakos}, A., {Nelson}, R.~H., \&
  {Zakrzewski}, B. 2014, \mnras, 437, 3718

\end{thebibliography}

\appendix
\section{MTGR coefficients for stars with $-1.0 \leqslant {\rm [M/H]} < -0.4$} 
\label{lowmetal}
\ \\

\noindent $a = -0.213170\,\mathrm{[M/H]}+0.641994$ (for $-0.5 \leqslant {\rm [Fe/H]} < -0.4$) \\
$a = +0.080416\,\mathrm{[M/H]}+0.788564$ (for $-1.0 \leqslant {\rm [Fe/H]} < -0.5$)

\noindent $b = +0.051362\,\mathrm{[M/H]}^{2} + 0.175915\,\mathrm{[M/H]}+0.402455$

\noindent $c = +0.500361\,\mathrm{[M/H]}^{3} +1.059979\,\mathrm{[M/H]}^{2} +
\\ \null\hspace{0.825cm} 0.447992\,\mathrm{[M/H]}+0.831917$

\noindent $d = -0.040907\,\mathrm{[M/H]}^{2} + 0.024522\,\mathrm{[M/H]}-0.201473$

\noindent $e = +0.215200\,\mathrm{[M/H]}-0.174482$ \\

\begin{table}[h]
\caption{Validity range of the MTGR for $-1.0 \leqslant {\rm [M/H]} < -0.4$.}
\centering
\begin{tabular}{llcc}
\hline\hline
$\rm{[M/H]}$ & Z & $T_{\rm eff}~[{\rm K}]$ range & log\,$g$ limit \\
\hline
$-1.0$ & $0.00155$ & $7950 - 26\,000$ & $\gtrsim 3.76$ \\
$-0.9$ & $0.00194$ & $7870 - 25\,500$ & $\gtrsim 3.73$ \\
$-0.8$ & $0.00244$ & $7670 - 25\,500$ & $\gtrsim 3.70$ \\
$-0.7$ & $0.00307$ & $7610 - 25\,500$ & $\gtrsim 3.67$ \\
$-0.6$ & $0.00385$ & $7430 - 24\,500$ & $\gtrsim 3.64$ \\
$-0.5$ & $0.00483$ & $7090 - 24\,000$ & $\gtrsim 3.61$ \\
\hline
\end{tabular} \label{tab:valid}
\end{table}

\begin{figure}
\begin{center}
\includegraphics[width=\columnwidth]{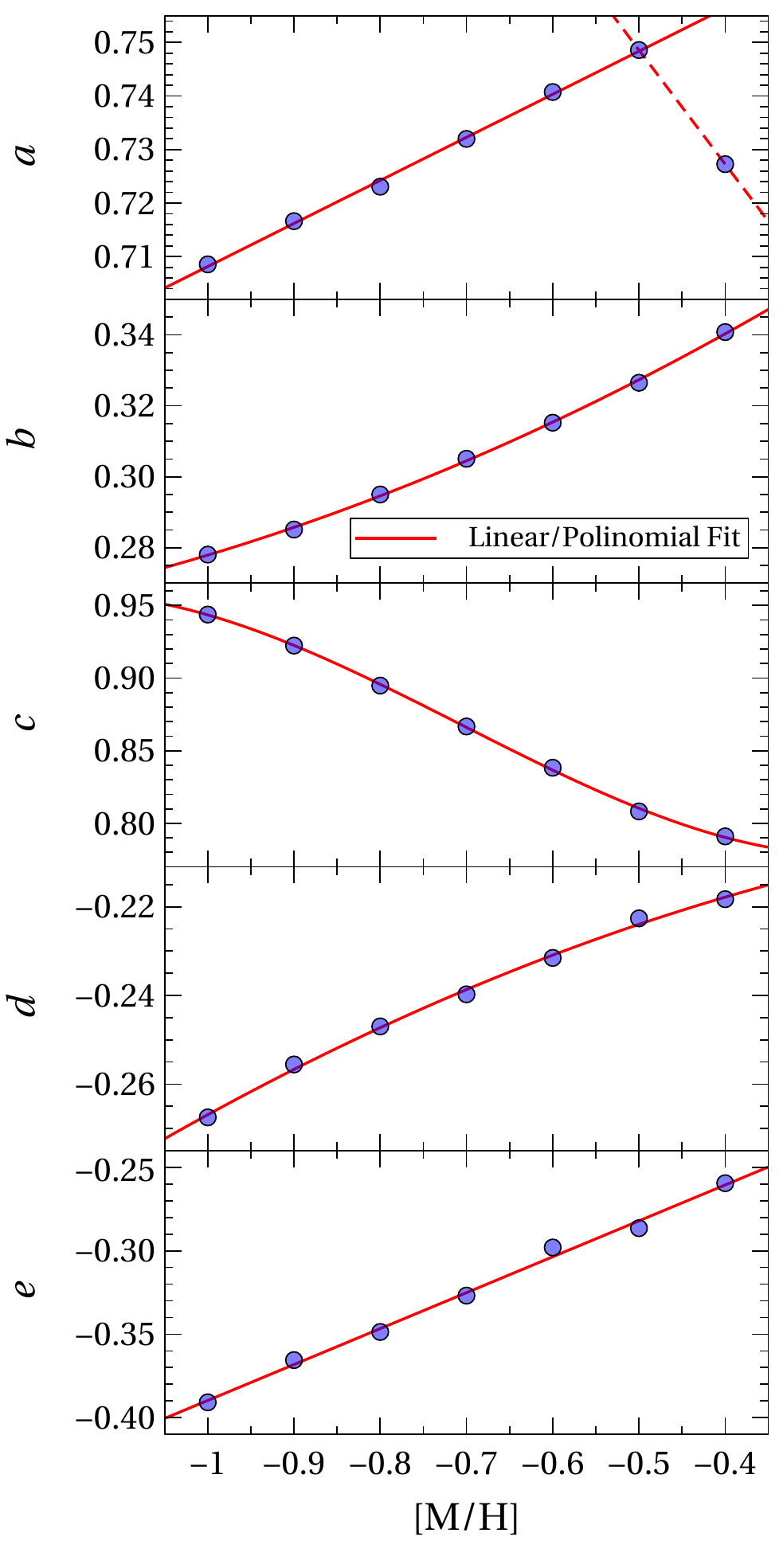}
\caption{Changes in the MTGR coefficients with metallicity in the range of $-1.0 \leqslant {\rm [M/H]} \leqslant -0.4$.}
\label{figA1}
\end{center}
\end{figure}

\onecolumn

\section{Regression analysis for the uncertainty expressions}

\begin{figure*}[!ht]
\begin{center}
\includegraphics[width=0.83\textwidth]{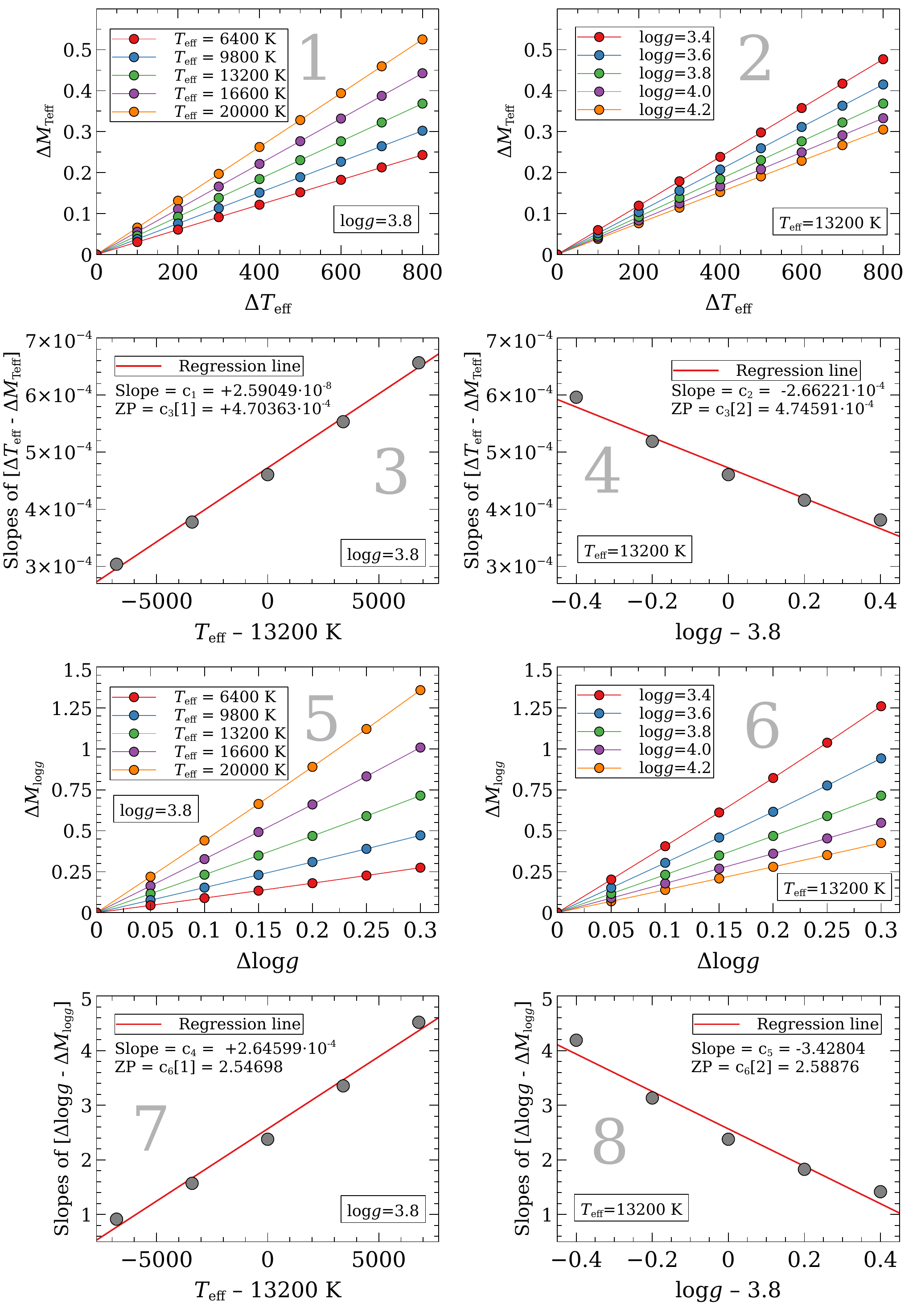}\\
\caption{Regression analysis and determining the coefficients in the uncertainty estimation formula for the MTGR with the metallicity-independent coefficients. The zero points (ZPs) $c_3$ and $c_6$ were calculated by averaging the individual calculations, i.e., $c_{3}=(c_{3}[1]+c_{3}[2])/2$ and $c_{6}=(c_{6}[1]+c_{6}[2])/2$. }
\label{FigB1}
\end{center}
\end{figure*}

\begin{figure*}[!ht]
\begin{center}
\includegraphics[width=0.85\textwidth]{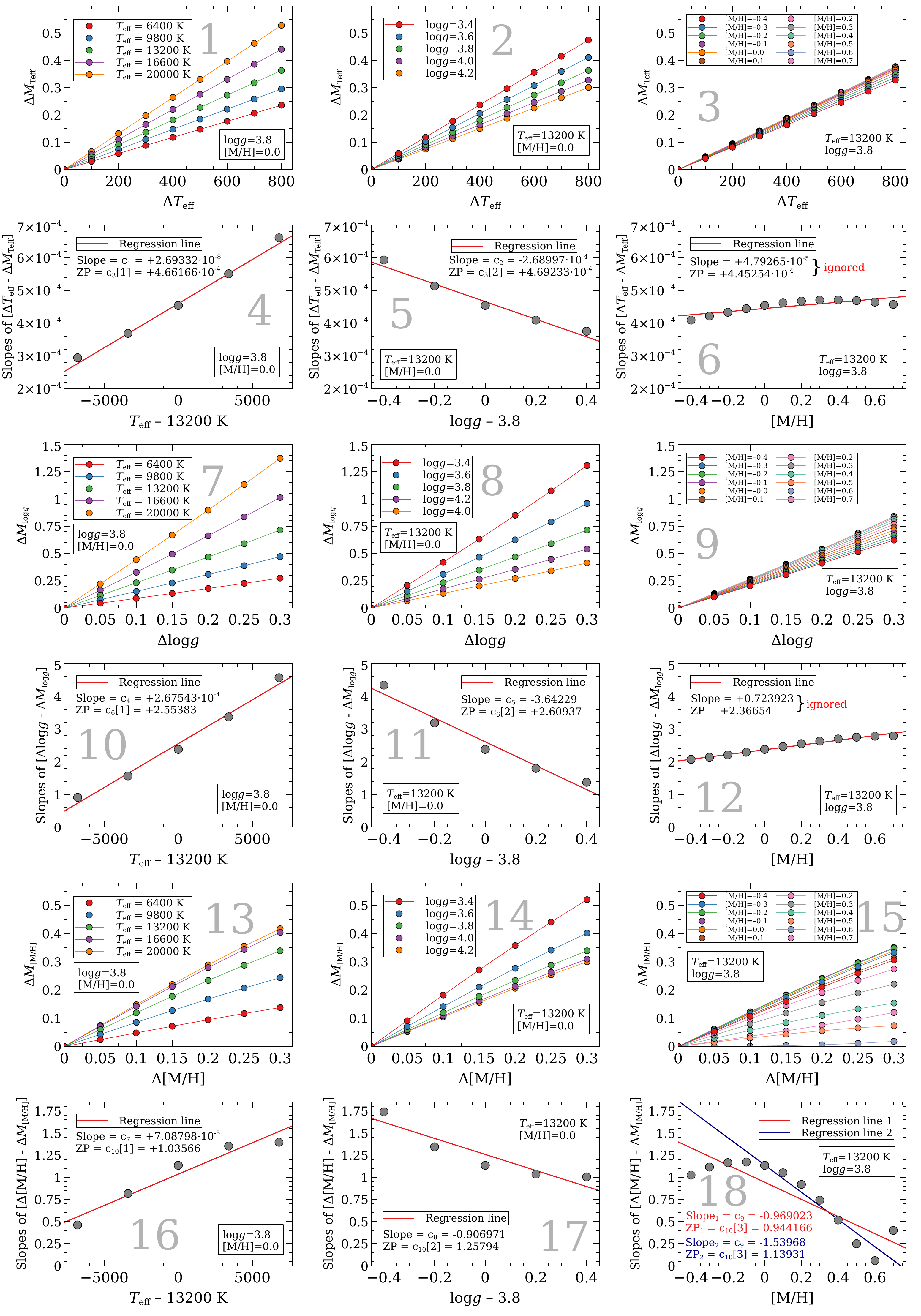}\\
\caption{Regression analysis and calculating the coefficients in the uncertainty estimation formula for the MTGR with the metallicity-dependent coefficients. The slopes in graphs 6 and 12 were too small and ignored (see text). Two solutions were provided for $c_{9}$ and $c_{10}$: the regression line 1 is for the entire [M/H] range and the regression line 2 is a better estimate for $-0.1 \leqslant [{\rm M/H}] \leqslant 0.6$. The zero points (ZPs) $c_{3}$, $c_{6}$ and $c_{10}$ were calculated by averaging the individual calculations, i.e., $c_{3}=(c_{3}[1]+c_{3}[2])/2$, $c_{6}=(c_{6}[1]+c_{6}[2])/2$ and $c_{10}=(c_{10}[1]+c_{10}[2]+c_{10}[3])/3$.}
\label{FigB2}
\end{center}
\end{figure*}

\twocolumn
\section{Uncertainty estimation formula including uncertainties in $[{\rm \alpha/Fe}]$}
\label{alphauncert}

$\Delta M \simeq (((c_{1} T_{\rm{0}}+c_{2} g_0+c_{3})\Delta T_{\rm eff})^2 + $

\ \ \ \ \ $((c_{4}T_{\rm{0}}+c_{5}g_{0}+c_{6})\Delta{\rm log}\,g)^2 + $

\ \ \ \ \ $((c_{7}T_{\rm{0}}+c_{8}g_{0}+c_{9}\rm{[M/H]+c_{10}})\Delta{\rm { [M/H]}})^2 + $

\ \ \ \ \ $(((c_{11}{\rm log}(M_{\rm MTGR})+c_{12})M_{\rm MTGR}\Delta[{\rm \alpha/Fe}])^2)^{1/2}$ \\

%M_{\rm MTGR}
%\Delta[{\rm \alpha/Fe}]

\noindent where: \\

$T_{\rm{0}} = T_{\rm{eff}} - 13\,200\,\rm{K}$

$g_{0} = {\rm log}\,g - 3.8 $ 

$\Delta[{\rm \alpha/Fe}]=$ uncertainty of $[{\rm \alpha/Fe}]$ 

$M_{\rm MTGR}$: mass calculated by the MTGR \\

$c_{1}= +2.69332\cdot10^{-8}$

$c_{2}= -2.68997\cdot10^{-4}$

$c_{3}= +4.67700\cdot10^{-4}$

$c_{4}= +2.67543\cdot10^{-4}$

$c_{5}= -3.64229$

$c_{6}= +2.58160$

$c_{7}= +7.08798\cdot10^{-5}$

$c_{8}= -0.906971$

$c_{9}= -0.969023$ (for the entire [M/H] range)

$c_{9}= -1.53968$ (a better estimate for $-0.1 \leqslant [{\rm M/H}] \leqslant 0.6$)

$c_{10}= +1.07926$ (for the entire [M/H] range)

$c_{10}= +1.14431$ (a better estimate for $-0.1 \leqslant [{\rm M/H}] \leqslant 0.6$) 

$c_{11}= k_1 = -0.17$

$c_{12}= k_2 = 0.255$ \\

\end{document}